\documentclass[12pt,a4paper]{article}
\usepackage[T2A]{fontenc}
\usepackage[cp866]{inputenc}
\usepackage[english]{babel}
\usepackage[dvips]{graphicx }
\usepackage{axodraw,epsfig}
\usepackage{amsmath,euscript,amssymb}
\newcommand{\be}{\begin{equation}}
\newcommand{\ee}{\end{equation}}
\newcommand{\bea}{\begin{eqnarray}}
\newcommand{\eea}{\end{eqnarray}}
\newcommand{\non}{\nonumber}

\title{Radiative Kaon Decay in Chiral Perturbation Theory}
\medskip
\author{A.Z. Dubni\v{c}kov\'{a}$^{1}$, S. Dubni\v{c}ka$^{2}$,
E. Goudzovski$^{3,4}$, \\
V.N. Pervushin$^{4}$, M. Se\v cansk\'y$^{2,4}$ \\
\\
$^{1}${\small Department of Theoretical Physics, Comenius
University, Bratislava, Slovak Republic}\\
$^{2}${\small  Institute of Physics, Slovak Academy of Sciences,
Bratislava, Slovak Republic}\\
$^{3}${\small Scuola Normale Superiore, I-56100 Pisa, Italy}\\
$^{4}${\small Joint Institute for Nuclear Research, 141980 Dubna,
Russia} }

\begin{document}

\maketitle

\begin{abstract}

We analyze the possibility of experimental investigation of new
low-energy relations between the values of resonance masses in the
meson form factors and the differential rate of radiative kaon decay
$K^+\to \pi^+ e^+e^-(\mu^+\mu^- )$ at the current level of the
experimental precision. A set of  arguments is listed in favour of
that these relations can be a  consequence of weak static
interactions in the Standard Model.

\end{abstract}

\vspace{-13cm}

\hfill Dedicated to the 70-th anniversary of professor M.K. Volkov

\vspace{13cm}


 \vspace{2cm}

{\large \centerline{The results were presented at}\vspace{.1cm}
\centerline{the 5th NA48 Mini-Workshop on kaon physics}\vspace{.1cm}
\centerline{(CERN, December 12, 2006).}}

\newpage

\section*{Introduction}

 The radiative kaon decay amplitudes $K^+\to \pi^+ e^+e^-(\mu^+\mu^- )$
 are  of great interest of the chiral perturbation theory
 (ChPT)~\cite{fpp,vp1,gs} because
  in the lowest order of ChPT the decay amplitudes are
 equal to zero~\cite{bvp,ecker,belkov01,ecker-r,da98}.
 There two opinions about the next order of ChPT.

 The first opinion is that in the next order of ChPT the
 baryon (or quark) loops dominate~\cite{bvp,belkov01}.
  As  these fermion loops
   also determine
 meson form factors in the low energy region \cite{fpp}, in this case, one can
  supposes that
   ChPT calculation~\cite{fpp,vp1,gs,bvp}
 points out possible relations between the low energy parameters
 of meson form factors and the differential rates of
 the radiative kaon decays. These relations can arise
  if we keep in the ChPT diagrams
 the real vector meson propagators and take into account
 the quantum numbers of the nearest resonances in
 possible vertices  \cite{05,117}.

 The second opinion  is
 that in ChPT the meson loops dominate.
   It was shown~\cite{ecker,ecker-r,da98}, in the framework of
    the accepted approach to Standard Model \cite{am,va,db}
    with the point-like approximation of weak
    interactions,
  that  these meson loops can completely destruct the meson form factor structure
   of the radiative kaon decay amplitudes in the low energy region.

   In this paper, we show that the situation with
   ChPT for radiative kaon decay amplitudes
   is  more complicate.
    There are two types of the meson loops. The first of them
    are provided by the normal ordering
    of the weak static interactions, and the second are retarded ones.

Recall
that static interactions arise in the Hamiltonian approach to the
Standard Model (SM) of electroweak (EW) interactions \cite{hpp,252}
in contrast to the conventional one~\cite{am,va} based on heuristic
Lorentz gauge formulation~\cite{fp1,db}, where the static
interactions are absent. These static interactions suppress any
retarded meson loop contributions \cite{ecker,ecker-r,da98} that can
destruct the meson form factor structure of the decay amplitudes.
The meson loop contributions can be  only the tadpole loop diagrams
following from the normal ordering of the static interaction. This
ordering  results in an effective action with $\triangle T=1/2$ rule
\cite{kp,CP-enhancement} with one unknown parameter $g_8$ that can
be fixed from other decays as  $g_8=5.1$. The dominance of weak
static interactions justifies the application of low energy chiral
perturbation theory~\cite{fpp,vp1,gs} as an efficient method of
description of kaon decay
processes~\cite{bvp,ecker,belkov01,CP-enhancement}.

In this paper we study the possibility of
 extracting information about the meson form factors
 from  the $K^+\to\pi^+e(\mu)^+e(\mu)^-$ processes at the current level
 of the experimental precision.

 The plan of the paper is the following. In Section 1
  we present the explicit expressions for
  amplitudes of the processes $K^+\to\pi^+l^+l^-$ in terms of meson
form factors. In Section 2 we discuss possibilities of the
corresponding experimental tests. Manifestations of the static
interactions in decay rates in SM are discussed in Section 3.

%
%
\section{Relations between form
factors and radiative $K$ decay amplitude in ChPT}

\subsection{Chiral bosonization of EW interaction}

 It is conventional to  describe  weak decays  in the framework of
 electroweak (EW) theory at the quark QCD level
 including  current vector boson weak interactions \cite{am,va}
 \be\label{ch111}
 \mathcal{L}_{(J)}=-(J^{-}_{\mu}W^{+}_{\mu}
 +J^{+}_{\mu}W^{-}_{\mu})=
 -\frac{e}{2\sqrt{2}\sin\theta_{W}}(\underline{J}^{-}_{\mu}W^{+}_{\mu}
 +\underline{J}^{+}_{\mu}W^{-}_{\mu}),
 \ee
 where
 $\underline{J}^{+}_{\mu}=\bar{d}'\gamma_{\mu}(1-\gamma_{5})u,$
 $\bar{d}'=d\cos\theta_{C}+s\sin\theta_{C}$, and
 $\theta_{C}$ is the Cabbibo angle ($\sin\theta_{C}=0.223$).

However, a consistent theory of QCD at large distances has not been
constructed yet. Therefore, the most efficient method
 of analysis in kaon decay physics
 \cite{bvp,ecker,belkov01,CP-enhancement} is the ChPT~\cite{vp1,gs}.
 The quark content of $\pi^{+}$ and $K^{+}$
 mesons $\pi^{+}=(\bar{d},u), K^{+}=(\bar{s},u),
  \overline{K}^{0}=(\bar{s},d) $  leads to the
 effective chiral hadron currents $\underline{J}^{\pm}_\mu$ in
 the  Lagrangian (\ref{ch111})
 \be\label{chl1}
 \underline{J}^{\pm}_{\mu}=[\underline{J}^1_{\mu}{\pm}i\underline{J}^2_{\mu}]\cos\theta_{C}\,+
 [\underline{J}^4_{\mu}{\pm}i\underline{J}^5_{\mu}]\,\sin\theta_{C}\,,
 \ee
 where using the Gell-Mann matrices $\lambda^k$ one can define
 the meson current as \cite{vp1}
 \be\label{c3k}
 i\sum\limits \lambda^k
 \underline{J}^k_{\mu}=i\lambda^k(V^{k}_{\mu}-A^{k}_{\mu})^{k}=F^2_\pi
 e^{i\xi}\partial_\mu e^{-i\xi},
 \ee
\be\label{c4}
 \xi=F_\pi^{-1}\sum\limits_{k=1}^{8}M^k\lambda^k=F_\pi^{-1}\left(%
\begin{array}{ccc}
  \pi^0+\dfrac{\eta}{\sqrt{3}} & \pi^+\sqrt{2} & K^+\sqrt{2} \\
  \pi^-\sqrt{2}  & -\pi^0+\dfrac{\eta}{\sqrt{3}} & K^0\sqrt{2} \\
  K^-\sqrt{2} & \overline{K}^0\sqrt{2} & -\dfrac{2\eta}{\sqrt{3}} \\
\end{array}%
\right).
 \ee
 In the first orders in mesons one can write

 \be \label{chl2}
 V^{-}_{\mu}=\sqrt{2}\,\,[\,\sin\theta_{C}\,
 (K^{-}\partial_{\mu}\pi^{0}-\pi^{0}\partial_{\mu}K^{-} )\,
 +\cos\theta_{C}\,(\pi^{-}\partial_{\mu}\pi^{0}-
 \pi^{0}\partial_{\mu}\pi^{-})\,]+...
  \ee
  and
  \be\label{chl3}
  A^{-}_{\mu}=\sqrt{2}\,F_{\pi}\,(
 \partial^{\mu}K^{-}\sin\theta_{C} +
 \partial^{\mu}\pi^-\cos\theta_{C})+...,
 \ee
here $F_{\pi}\simeq 92.4$ MeV. The right form of the chiral
Lagrangian of the electromagnetic interaction of mesons  can be
constructed by the covariant derivative
 $
 \partial_{\mu}\chi^{\pm}\to D_{\mu}\chi^{\pm}
\equiv(\partial_{\mu}\pm ieA_{\mu})\chi^{\pm},
 $
 where $\chi^{\pm}=K^{\pm},\pi^{\pm}$.

We suppose also that the quark content of the mesons determines
hadronization of QCD \cite{5,6} conserving its chiral and gauge
symmetries.

\subsection{The $K^+\to\pi^+l^+l^-$ amplitude}

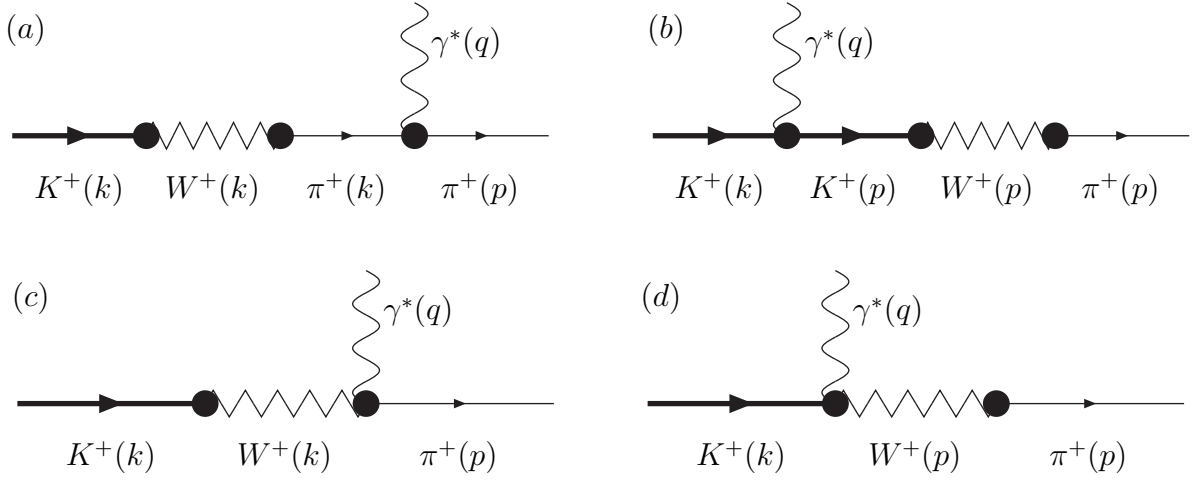
\begin{figure}
\centering
\begin{minipage}[c]{0.45\hsize}
\begin{picture}(200,100)(0,0)\Vertex(150,50){5}\Vertex(50,50){5}\Vertex(100,50){5}
\ZigZag(50,50)(100,50){5}{5}\ArrowLine(100,50)(150,50)\ArrowLine(150,50)(200,50)
\Photon(150,50)(150,100){5}{3} \SetWidth{2.0}
\ArrowLine(0,50)(50,50)
\Text(5,90)[]{$(a)$}\Text(25,30)[]{$K^{+}(k)$}
\Text(125,30)[]{$\pi^{+}(k)$}\Text(175,30)[]{$\pi^{+}(p)$}\Text(75,30)[]{$W^{+}(k)$}
\Text(170,85)[]{$\gamma^{*}(q)$}
\end{picture}
\end{minipage}
\hspace*{5mm}
\begin{minipage}[c]{0.45\hsize}
\begin{picture}(200,100)(0,0)\Vertex(150,50){5}\Vertex(50,50){5}\Vertex(100,50){5}
\ZigZag(100,50)(150,50){5}{5}\ArrowLine(150,50)(200,50)
\Photon(50,50)(50,100){5}{3} \SetWidth{2.0}
\ArrowLine(0,50)(50,50)\ArrowLine(50,50)(100,50)\Text(5,90)[]{$(b)$}
\Text(25,30)[]{$K^{+}(k)$}
\Text(125,30)[]{$W^{+}(p)$}\Text(175,30)[]{$\pi^{+}(p)$}\Text(75,30)[]{$K^{+}(p)$}
\Text(70,85)[]{$\gamma^{*}(q)$}
\end{picture}
\end{minipage}
\begin{minipage}[c]{0.45\hsize}
\begin{picture}(200,100)(0,0)
\Vertex(70,50){5}\Vertex(130,50){5}
\ZigZag(70,50)(130,50){5}{5}\ArrowLine(130,50)(200,50)
\Photon(130,50)(130,100){5}{3} \SetWidth{2.0}
\ArrowLine(0,50)(70,50)
\Text(5,90)[]{$(c)$}\Text(35,30)[]{$K^{+}(k)$}
\Text(100,30)[]{$W^{+}(k)$}\Text(165,30)[]{$\pi^{+}(p)$}
\Text(150,85)[]{$\gamma^{*}(q)$}
\end{picture}
\end{minipage}\hspace*{5mm}
\begin{minipage}[c]{0.45\hsize}
\begin{picture}(200,100)(0,0)
\Vertex(70,50){5}\Vertex(130,50){5}
\ZigZag(70,50)(130,50){5}{5}\ArrowLine(130,50)(200,50)
\Photon(70,50)(70,100){5}{3} \SetWidth{2.0} \ArrowLine(0,50)(70,50)
\Text(5,90)[]{$(d)$}\Text(35,30)[]{$K^{+}(k)$}
\Text(100,30)[]{$W^{+}(p)$}\Text(165,30)[]{$\pi^{+}(p)$}
\Text(90,85)[]{$\gamma^{*}(q)$}
\end{picture}
\end{minipage}
\caption{ $K^+\to\pi^+\gamma^*$ diagrams for effective Lagrangian.}
\label{ac}\end{figure}

The result of calculation of the amplitude of the process
$K^+\to\pi^+l^+l^-$ ($l=e;\mu$) in the framework of the chiral
Lagrangian (\ref{ch111})--(\ref{chl3}) including phenomenological
meson form factors denoted by fat dots in Fig.~\ref{ac} takes the
form

 \be\label{ampl}
  T_{(K^+\to\pi^+l^+l^-)}= 2g_8eG_{\rm EW}L_{\nu}
 D^{\gamma(rad)}_{\mu\nu}(q)(k_{\mu }+p_{\mu })\,\,\,
 {\cal T}(q^2,k^2,p^2),
 \ee
 where $g_8=5.1$ is the effective enhancement coefficient
 \cite{ecker,kp},
\be\label{g}
  G_{\rm EW}= \frac{\sin\theta_{C} \cos\theta_{C}}{8
  M^{2}_{W}}\frac{e^{2}}{\sin^{2}\theta_{W}}\equiv
  \sin\theta_{C} \cos\theta_{C}\frac{G_F}{\sqrt{2}},
 \ee
  is the coupling constant,
  $L_{\mu}=\bar{l}\gamma_{\mu}l$ is leptonic current,
 \be\label{tp}
 {\cal T}(q^2,k^2,p^2)=F^2_{\pi}
 \left[\frac{f^{V}_{\pi}(q^{2})k^{2}}{m^{2}_{\pi}-k^{2}-
 i\epsilon}+\frac{f^{V}_{K}(q^{2})p^{2}}{M^{2}_{K}-p^{2}-i\epsilon}+
 \frac{f^{A}_{K}(q^{2})+f^{A}_{\pi}(q^{2})}{2}\right],
 \ee
 and
 $f^{(A,V)}_{\pi,K}(q^{2})$ are meson form factors.
 On the mass-shell the sum (\ref{tp}) takes the form
 \bea\label{t0}
 {\cal T}(q^2,M_K^2,m_\pi^2)&=&\\\nonumber
 =~{\cal T}(q^2)&=&F^2_\pi  \Bigg[\frac{f^{A}_{K}(q^{2})+f^{A}_{\pi}(q^{2})}{2}
 - f^{V}_{\pi}(q^{2})+
 \Big[f^{V}_{K}(q^{2})- f^{V}_{\pi}(q^{2})\Big]\frac{m_\pi^2}{M_K^2-m_\pi^2}
  \Bigg].
 \eea
The $K^+\to\pi^+l^+l^- $ amplitude vanishes at tree
level~\cite{bvp,ecker}, where the form factors are equal to unity:
${\cal T}(q^2)|_{f^{V}=f^{A}=1}=0$.

In terms of the two standard Dalitz plot variables $q^2$ and $t^2$
representing the squares of invariant masses of $l^+l^-$ and
$\pi^+l^+$ pairs, respectively, the amplitude (\ref{t0}) leads to
the following decay rate for the transition $K^+\to\pi^+l^+l^-$:
\begin{equation}
\Gamma(q^2,t^2) = C \int\limits_{4m^2_l}^{(M_K-m_\pi)^2}dq^2\cdot
|F(q^2)|^2 \int\limits_{t_{min}^2(q^2)}^{t_{max}^2(q^2)}dt^2 \cdot
\eta(q^2,t^2), \label{eq:2kinvar}
\end{equation}
where
\begin{eqnarray}
\eta(q^2,t^2) &=&
(2t^2+q^2-2m_\pi^2-2m_l^2)(2M_K^2+2m_l^2-2t^2-q^2)+\nonumber\\
&&+q^2(q^2-2M_K^2-2m_\pi^2)
\end{eqnarray}
and
 \bea\label{f3}\nonumber
F(q^2)&=&\dfrac{(4\pi)^2{\cal T}(q^2)}{q^2}=\\\label{1f3}
&=&\frac{(4\pi F_\pi)^2}{q^2}
  \left[\frac{f^{A}_{K}(q^{2})+f^{A}_{\pi}(q^{2})}{2}
 - f^{V}_{\pi}(q^{2})+ \Big[f^{V}_{K}(q^{2})-
 f^{V}_{\pi}(q^{2})\Big]\frac{m_\pi^2}{M_K^2-m_\pi^2}
  \right].
 \eea
The $t^2$-dependence does not contain information about the
combination of the form factors $F(q^2)$, which is of our interest.
Integration of (\ref{eq:2kinvar}) over $t^2$ yields
 \be\label{f2}
 \Gamma(q^2)=C
 \int\limits_{4m^2_l}^{(M_K-m_\pi)^2}
 {\frac{d q^2}{M_K^2} \rho(q^2)}|F(q^2)|^2.
 \ee
Here (see \cite{ecker})
 \bea\label{gf3}
 C=\frac{(s_1c_1c_3)^2g_8^2G^2_F}{(4\pi)^4}
 \frac{\alpha^2 M_K^5}{24\pi}\Big|_{g_8=5.1}=1.37 \times 10^{-22}~\mbox{\rm GeV},
 \eea
\bea\label{rho}\nonumber
 \rho(q^2)&=&\left(1-\frac{4m_l^2}{q^2}\right)^{1/2}
 \left(1+\frac{2m_l^2}{q^2}\right)~\!
 \lambda^{3/2}
 (1,q^2/M^2_K,m^2_\pi/M^2_K),\\\nonumber
\lambda (a,b,c)&=&a^2+b^2+c^2-2(ab+bc+ca), \eea and $s_1 c_1 c_3$ is
the product of Cabibbo-Kobayashi-Maskawa matrix elements
$V_{ud}V_{us}$.

\section{Parameterization of $F(q^2)$}

\subsection{Parameterization with meson loops}

It follows from (\ref{f3}), taking into account
$f_{\pi}^{V}(q^2)\simeq f_K^{V}(q^2)$ and $f_{\pi}^{A}(q^2)\simeq
f_{K}^{A}(q^2)$:
\begin{equation}
F(q^2)=\frac{(4\pi F_\pi)^2}{q^2}
  \left[f^{A}(q^{2}) - f^{V}(q^{2})\right].
\label{Fsimple}
\end{equation}
We discuss the differential $K^+\to\pi^+l^+l^-$ decay rate
(\ref{eq:2kinvar}, \ref{f2}) in the ChPT~\cite{fpp,vp1} with pion
and baryon loop contributions leading to meson form
factors~\cite{fpp,bvp}:
 \be\label{1vform}
 \begin{array}{ccl}
 f^{V}(q^2)&=&1+M^{-2}_\rho q^2+
 \alpha_{0}\Pi_\pi(q^2)+...~;\\
 f^{A}(q^2)&=&1+M^{-2}_a
 q^2+...
 \end{array}
 \ee
We parameterize the terms linear in $q^2$ (determined by the baryon
and meson loops~\cite{fpp,bvp,belkov01,CP-enhancement}) by the
values of resonance masses~\cite{7} $M_\rho=775.8$ MeV,
$I^G(J^{PC})=1^+(1^{--})$ and $M_a=984.7$ MeV,
$I^G(J^{PC})=1^-(0^{++})$, \be\label{m1}
\alpha_0=\dfrac{4}{3}\cdot\dfrac{m_\pi^2}{(4\pi F_\pi)^2}
=0.01926,\ee and the nonlinear term of the pion loop
contribution~\cite{fpp,vp1} is given by
\begin{equation}
\begin{array}{rcll}
\Pi_\pi(t) &=& (1-\bar t)\left(\dfrac{1}{\bar t}-1\right)^{1/2}
\arctan\left(\dfrac{\bar t^{1/2}}{(1-\bar t)^{1/2}}\right)-1,&
\bar t=\dfrac{t}{(2m_\pi)^2}<1;\\
\Pi_\pi(t) &=& \dfrac{\bar t-1}{2}\left(1-\dfrac{1}{\bar
t}\right)^{1/2}\left\{i\pi- \log \dfrac{\bar t^{1/2}+(\bar
t-1)^{1/2}}{\bar t^{1/2}-(\bar t-1)^{1/2}}\right\}-1,& \bar t\geq 1.
\end{array}
\end{equation}
In order to introduce the resonant behavior of the form factors, the
following Pad\'e-type approximations \cite{pade} to the expressions
(\ref{1vform}) are considered:
\begin{equation}
\label{fmodi}
\begin{array}{rcl}
f_1^{V}(q^2) &=& \gamma\left[1 - \{ M^{-2}_\rho q^2 + \alpha_0\cdot
\Pi_\pi(q^2)\} /\gamma\right]^{-1} + (1-\gamma),\\
f_1^{A}(q^2) &=& \left(1-M_a^{-2}q^2\right)^{-1}.
\end{array}
\end{equation}
Here the parameter $\gamma=0.85$ effectively accounts for higher
order loops, and is chosen such as to put the position of maximum of
$f_1^V(q^2)$ to $q^2=M_\rho^2$.

\subsection{Predictions for integrated and differential decay rates}

The form factors (\ref{fmodi}) lead to the following decay branching
ratios and muon/electron ratio $R={\rm
Br}(K^+\to\pi^+\mu^+\mu^-)/{\rm Br}(K^+\to\pi^+e^+e^-)$:
$$
{\rm Br}(K^+\to\pi^+e^+e^-) = 3.88 \times 10^{-7},~~ {\rm
Br}(K^+\to\pi^+\mu^+\mu^-) = 1.23 \times 10^{-7},~~
 R = 0.318.
$$
These branching fractions are highly sensitive to the values of
$M_a$ and $M_\rho$ used in the parameterization:
\begin{equation}
\begin{array}{ll}
\Delta{\rm Br}(ee)/{\rm Br}(ee)\approx 12(\Delta M_a/M_a),&
\Delta{\rm Br}(\mu\mu)/{\rm Br}(\mu\mu)\approx 10(\Delta
M_a/M_a),\\
\Delta{\rm Br}(ee)/{\rm Br}(ee)\approx-20(\Delta M_\rho/M_\rho),&
\Delta{\rm Br}(\mu\mu)/{\rm Br}(\mu\mu)\approx-17(\Delta
M_\rho/M_\rho).
\end{array}
\end{equation}
However this sensitivity largely cancels in the muon/electron ratio:
$$
\Delta R/R\approx -2(\Delta M_a/M_a),~~~ \Delta R/R\approx 2(\Delta
M_\rho/M_\rho).
$$
The sensitivity to the parameter $\gamma$ is smaller than to the
resonance masses:
$$
\Delta{\rm Br}(ee)/{\rm Br}(ee)\approx-0.5(\Delta\gamma/\gamma),~~
\Delta{\rm Br}(\mu\mu)/{\rm
Br}(\mu\mu)\approx-0.9(\Delta\gamma/\gamma),~~ \Delta
R/R\approx-0.4(\Delta\gamma/\gamma).
$$
Taking into account that the relative uncertainties of resonance
masses $M_a$ and $M_\rho$ that are about 1\%, our predictions can be
roughly quantified as follows:
\begin{eqnarray}
{\rm Br}(K^+\to\pi^+e^+e^-) &=& (3.9\pm0.8) \times 10^{-7},\nonumber\\
{\rm Br}(K^+\to\pi^+\mu^+\mu^-) &=& (1.2\pm0.3) \times 10^{-7},\nonumber\\
R &=& 0.32\pm0.01.\nonumber
\end{eqnarray}

\begin{figure}[tb]
\begin{center}
{\resizebox*{0.49\textwidth}{!}{\includegraphics{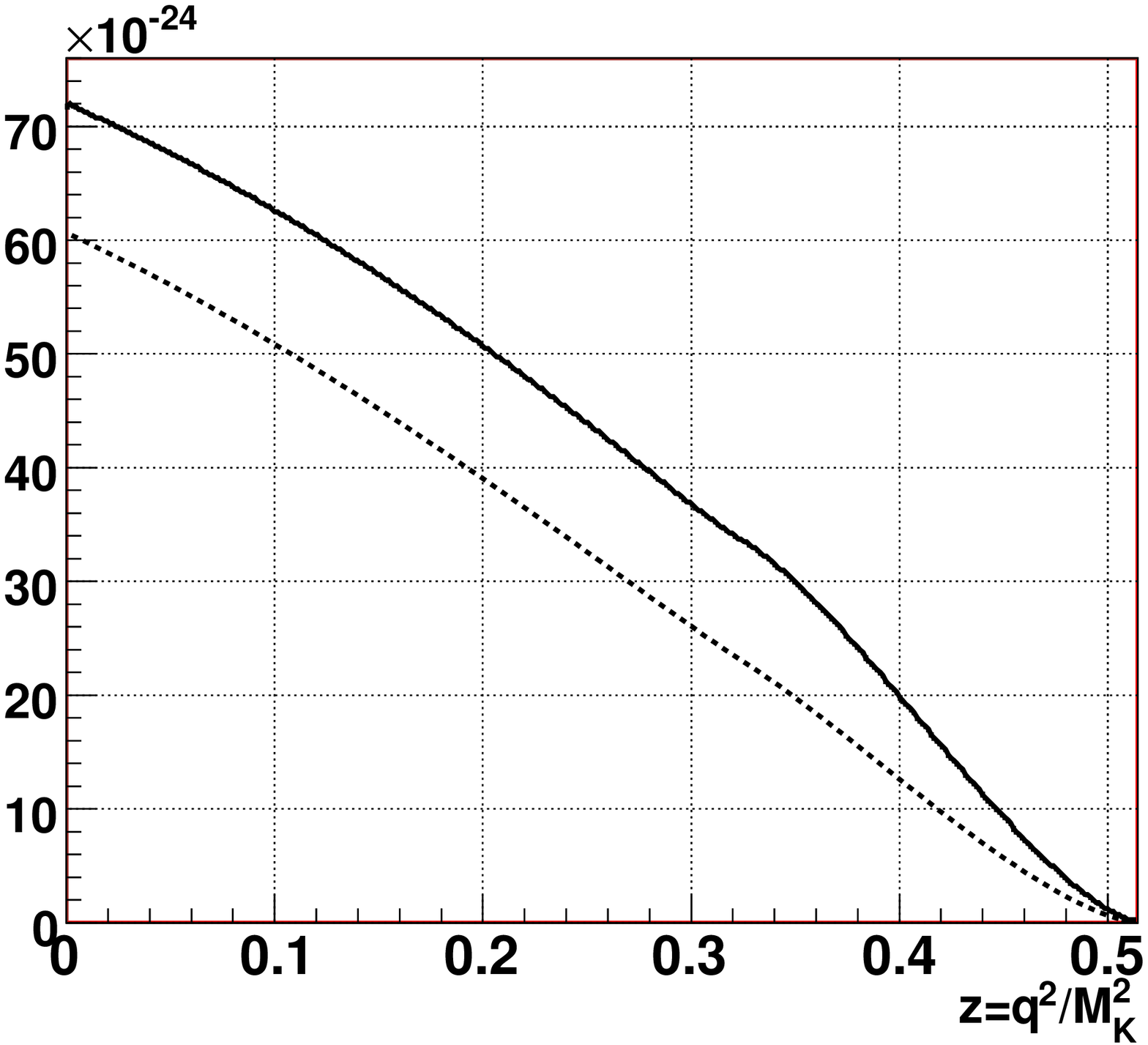}}}
{\resizebox*{0.49\textwidth}{!}{\includegraphics{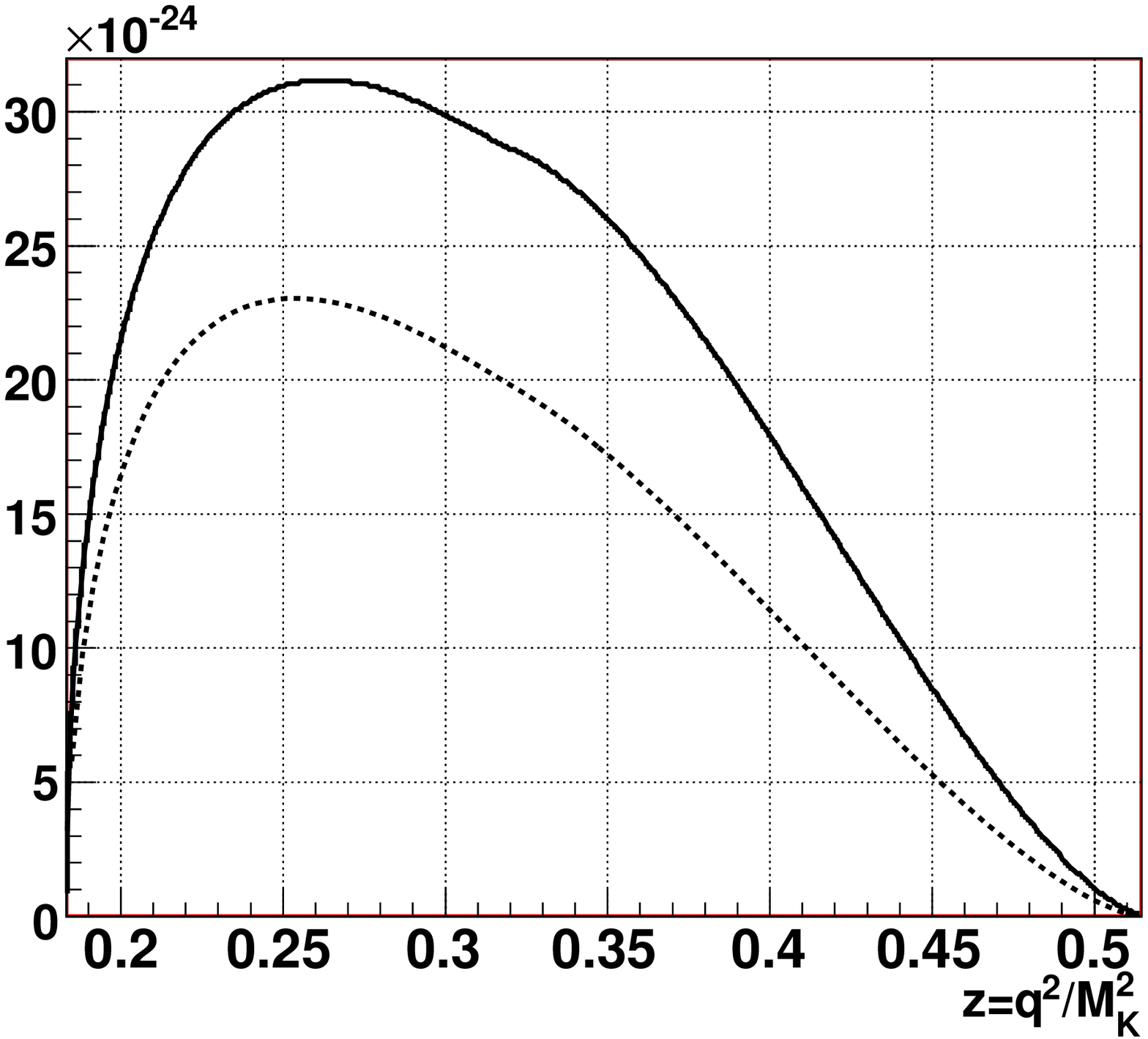}}}
\put(-270,166){\bf \large a} \put(-30,166){\bf \large b}
\end{center}
\vspace{-9mm} \caption{Differential rates of (a) $K^+\to\pi^+e^+e^-$
and (b) $K^+\to\pi^+\mu^+\mu^-$ decays as functions of
$z=q^2/M_K^2$. Solid lines: $F(q^2)$ parameterization (\ref{fmodi});
dotted lines: differential distribution measured for
$K^+\to\pi^+e^+e^-$, $z>0.1$ by~\cite{ap99} and extrapolated for
$z<0.1$ and for $K^+\to\pi^+\mu^+\mu^-$ using a model~\cite{da98}.
Given the large sensitivity of our calculation to values of $M_\rho$
and $M_a$, the calculation agrees with the experimental data.}
\label{fig:diffrates}
\end{figure}

Differential rates of $K^\pm\to\pi^\pm e^+e^-$ and $K^\pm\to\pi^\pm
\mu^+\mu^-$ decays corresponding to the parameterization
(\ref{fmodi}) are presented in Fig.~\ref{fig:diffrates} along with
the rates calculated extrapolating the available experimental data
on $K^\pm\to\pi^\pm e^+e^-$ decay~\cite{ap99} using a
model~\cite{da98}. Note that the experimentally accessible kinematic
region of the $K^+\to\pi^+e^+e^-$ decay is limited by a condition
$z=q^2/M_K^2\gtrsim(M_{\pi^0}/M_K)^2\approx 0.08$, while for the
$K^\pm\to\pi^\pm \mu^+\mu^-$ decay the whole kinematic range is
accessible.

\begin{figure}[tb]
\begin{center}
{\resizebox*{0.49\textwidth}{!}{\includegraphics{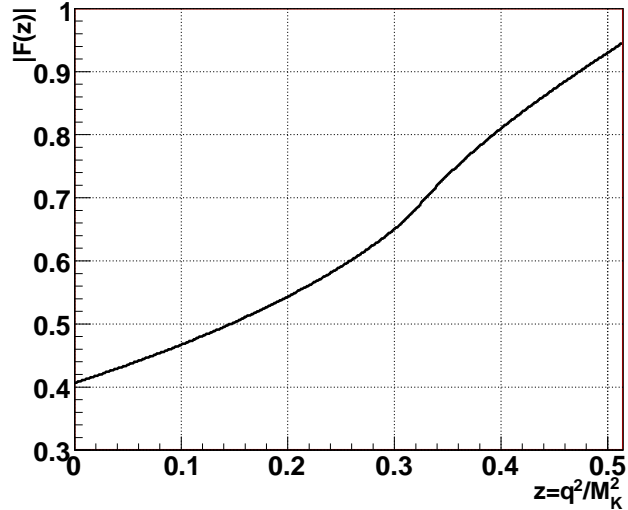}}}
\end{center}
\vspace{-9mm} \caption{The function $|F(q^2)|$ determined by the
relations (\ref{Fsimple}), (\ref{fmodi}).} \label{fig:F-DP}
\end{figure}

The function $|F(q^2)|$ determined by the relations (\ref{Fsimple}),
(\ref{fmodi}) is presented in Fig.\ref{fig:F-DP}. Its shape, being
approximated in terms of a linear form factor
$F(q^2)=F_0(1+\lambda\cdot q^2/M_K^2)$, leads to a form factor
varying from a minimum of $\lambda=1.4$ for $z=q^2/M_K^2=0$ to a
maximum of $\lambda=4.9$ for a point $z=q^2/M_K^2=0.32$
corresponding to $q^2\approx M_{\pi}^2$. An effective average form
factor slope of the $K^\pm\to\pi^\pm e^+e^-$ decay as would be
measured by an experiment in the accessible kinematic region
$q^2>M_{\pi^0}^2$ is estimated to be $\lambda\approx 2.3$. This
value should be subject to variation among different experiments,
depending, in particular, on the experimental acceptance as a
function of $q^2$.

\subsection{Prospects for experimental tests}

The experimental data for branching fractions, their ratio, and the
form factor slope (not considering a slope measurement in the
$K^\pm\to\pi^\pm\mu^+\mu^-$ channel which is subject to large
uncertainties) are~\cite{7}:
\begin{eqnarray}
{\rm Br}(K^+\to\pi^+e^+e^-) &=& (2.88\pm0.13) \times 10^{-7},\nonumber\\
{\rm Br}(K^+\to\pi^+\mu^+\mu^-) &=& (0.81\pm0.14) \times 10^{-7},\nonumber\\
R &=& 0.281\pm0.050,\nonumber\\
\lambda &=& 2.14\pm0.20.\nonumber
\end{eqnarray}
The experimental precision of ${\rm Br}(K^+\to\pi^+e^+e^-)$ is
mainly determined by a single measurement~\cite{ap99}. The
experimental uncertainty of ${\rm Br}(K^+\to\pi^+\mu^+\mu^-)$ is
dominated by a PDG error scale factor~\cite{7} emerging from
inconsistency of three measurements~\cite{ad97,ma00,pa02}.
Significant experimental improvements are expected in near future,
when the data sample collected by the NA48/2 experiment at CERN is
analyzed.

Our predictions for branching fractions of the two decays, their
ratio $R$, and the effective form factor slope are in agreement with
the experimental data. On the contrary, meson dominance
models~\cite{li99} fail to describe the effective form factor slope,
predicting substantially lower slope values.

Monte Carlo simulations involving realistic estimations of
experimental conditions show that a deviation of $K^\pm\to\pi^\pm
e^+e^-$ event distribution predicted by (\ref{fmodi}) from a
distribution corresponding to a linear form factor $F(q^2)$ (i.e.
the predicted dependence of the effective form factor slope
$\lambda$ on $q^2$) can be experimentally detected with a sample of
$\sim2\cdot10^{4}$ reconstructed decays, which is not far from the
capabilities of the present experiments in terms of kaon flux.


\section{Weak static interaction as the origin of  enhancement}
We listed the set of experimental arguments in favour of that a
relation between the form factors and radiative $K$ decay amplitude
takes place. In the following part of the paper we would like to
show that this relation is not occasional from theoretical point of view.

In any case, we are trying to address the questions: What are
contributions of other loop diagrams? What is the origin of the
enhancement coefficient $g_8$ in the amplitude (\ref{ampl})?
 What is origin of the coincidence of the
 resonance parameters  with the kaon decay ones?

 We show below that a reply to all these questions can be
the weak static interaction as a consequence of Dirac like radiation
variables in SM.

\subsection{The Radiation Variables in Standard Model}

 As it was shown by Dirac in
 QED \cite{dir}, the  static interactions in  gauge theories are an
 inevitable
 consequence of the general principles of QFT, including the vacuum
 postulate. In order to obtain a physical vacuum as a state with
 minimal energy, Dirac eliminated all zero momentum fields (with
 their possible negative contributions into the energy of the system) by
 solving the Gauss constraint and  dressing charged fields by the
 phase factors (see also \cite{sch2,f1,ni}).
 This elimination   leads to
 the {\it radiation variables} and the static interactions in both
 QED and the massive vector boson theory \cite{hpp}.
%

 In particular, in QED the radiation
 gauge-invariant  variables
$A^{(\rm
R)}_\mu(A)=A_\mu-\partial_\mu\dfrac{1}{\triangle}(\partial_k A^k)$
have   propagators $
\widetilde{J}^+_\mu D^{\rm R}_{\mu\nu}(q) \widetilde{J}^-_\nu =
\dfrac{\widetilde{J}^+_0\widetilde{J}^-_0}{\vec{q}^2} +
\left(\delta_{ij}-\dfrac{q_i q_j}{\vec{q}^2}\right)
\dfrac{\widetilde{J}^+_i\widetilde{J}^-_j}{q^2} $; 
while the Lorentz ones $A^{(\rm
L)}_\mu(A)=A_\mu-\partial_\mu\dfrac{1}{\Box}(\partial_\nu
A^\nu)$ \cite{db} have propagators 
$
 \widetilde{J}_\mu^+D^{L}_{\mu\nu}(q)\widetilde{J}_\nu^-=
-\widetilde{J}_\mu^+\dfrac{1}{q^2}\left(g_{\mu\nu}-\dfrac{q_\mu
q_\nu}{\Box} \right)\widetilde{J}_\nu^- $.

In order to demonstrate the inequivalence between the radiation
variables
  and the Lorentz ones, let us consider
 the electron-positron scattering amplitude
$T^R=\langle e^+,e^-|\hat S|e^+,e^-\rangle$. One can see that the
Feynman rules in the radiation gauge give the amplitude in terms of
the current $j_\nu=\!\bar e \gamma_\nu e$
 \bea\label{1wr}
 T^R=\frac{J^2_0}{\mathbf{q}^2}+
 \left(\delta_{ik}-\dfrac{q_iq_k}{\mathbf{q}^2}\right)
 \frac{J_iJ_k}{q^2+i\varepsilon}\equiv\frac{-J^2}{q^2+i\varepsilon}+
 \fbox{$\dfrac{(q_0J_0)^2-
 (\mathbf{q}\cdot\mathbf{j})^2}{\mathbf{q}^2[q^2+i\varepsilon]}$}~.
 \eea
 This amplitude coincides with the Lorentz gauge one,
 \be
 \label{2wr}
 T^L =
 -\frac{1}{q^2+i\varepsilon}
 \left[J^2-\fbox{$\dfrac{(q_0J_0-
 \mathbf{q}\cdot\mathbf{J})^2}{q^2+i\varepsilon}$}\,\,\right]~,
 \ee
 when the box terms  in Eq. (\ref{1wr}) can be
 eliminated. Thus, the Faddeev equivalence theorem \cite{f1} is valid,
  if the currents
 are conserved
 \be \label{3wr}
 q_0{J}_0-\mathbf{q}\cdot\mathbf{J}=qJ=0,
 \ee
%
 and the box terms
 are eliminated. It just the case, when
 the R variables  are equivalent to the L ones \cite{f1}.
 However, if elementary particles are off their mass-shell
 (in particular, in bound states) the currents are not
 conserved\footnote{The change of variables R $\to$ L  means a change of physical sources.
 In this case, the off  mass-shell L variable propagators
 lose the Coulomb pole forming
 the Coulomb atoms. The loss of the pole does not mean violation
 of the gauge invariance,
 because both the variables (R and L) can be defined as
 the gauge-invariant functionals of the initial gauge fields.}.


Radiation variables have vacuum as a state with the minimal energy,
whereas the Lorentz ones lose the vacuum postulate as the time
component give the negative contribution into the energy. Therefore,
  Schwinger  in \cite{sch2}
 {\it ... rejected all Lorentz
 gauge formulations as unsuited to the role of providing the fundamental operator
 quantization ...}

 Let we believe Schwinger, and consider the massive vector
 Lagrangian
\bea\nonumber
  {\cal L}=-\frac{1}{2}(\partial_\mu W_{\nu}^{+}&-&\partial_\nu
W_{\mu}^{+})(\partial_\mu W_{\nu}^{-}-\partial_\nu W_{\mu}^{-})+
M^{2}_{W}W_{\mu}^{+}W_{\mu}^{-}+\\&+&\left[J^{-}_\mu
W_\mu^++J^{+}_\mu W_\mu^-\right]\frac{e}{2\sqrt{2}\sin \theta_W}
\nonumber \eea
 in terms of radiation variables \cite{hpp}
$W^{\pm R}_\mu=W^{\pm}_\mu+ \partial_\mu 
[{1}/{(M_W^2-\triangle)}]\partial_kW^{\pm}_k$

In this case, instead of the standard propagator \cite{db}
\begin{equation} \label{3-5-4b}
 \widetilde{J}_\mu^+D^{L}_{\mu\nu}(q)\widetilde{J}_\nu^-=
-\widetilde{J}_\mu^+\frac{1}{q^2-M_W^2}\left(g_{\mu\nu}-\frac{q_\mu
q_\nu}{M_W^2} \right)\widetilde{J}_\nu^-
\end{equation}
we have the radiation one \cite{hpp}
\begin{equation} \label{3-5-4}
\widetilde{J}^+_\mu D^{\rm R}_{\mu\nu}(q) \widetilde{J}^-_\nu =
 \fbox{$\dfrac{\widetilde{J}^+_0\widetilde{J}^-_0}{\vec{q}^2+M_W^2}$}
+ \left(\delta_{ij}-\frac{q_i q_j}{\vec{q}^2+M_W^2}\right)
\frac{\widetilde{J}^+_i\widetilde{J}^-_j}{q^2-M_W^2}
\end{equation}
The R--propagator
 is regular in the limit $M_W\rightarrow 0$ and
 is
well behaved for large momenta. In the following we compare two
propagators $  D_{\mu\nu}^{L}$ and $D_{\mu\nu}^{R} $.

\subsection{Weak static interaction as the origin of  enhancement}

\begin{figure}
\centering
\begin{minipage}[c]{0.45\hsize}
\begin{picture}(200,100)(0,0)
\Vertex(70,50){5}\Vertex(130,50){5}
\ZigZag(70,50)(130,50){5}{5}\ArrowLine(130,50)(200,50)
\SetWidth{2.0} \ArrowLine(0,50)(70,50)
\Text(5,90)[]{$(a)$}\Text(35,30)[]{$K^{+}(k)$}
\Text(100,30)[]{$W^{+}(k)$}\Text(165,30)[]{$\pi^{+}(k)$}
\end{picture}
\end{minipage}\hspace*{5mm}
\begin{minipage}[c]{0.45\hsize}
\begin{picture}(200,100)(0,0)
\Vertex(70,50){5}\Vertex(130,50){5}\CArc(100,50)(30,0,180)
\ZigZag(70,50)(130,50){5}{5}\ArrowLine(130,50)(200,50)
\SetWidth{2.0} \ArrowLine(0,50)(70,50)
\Text(5,90)[]{$(b)$}\Text(35,30)[]{$K^{+}(k)$}\Text(95,90)[]{$\pi^{0}(k+l)$}
\Text(100,30)[]{$W^{+}(-l)$}\Text(165,30)[]{$\pi^{+}(k)$}
\end{picture}\end{minipage}
\caption{Axial (a) and vector (b) current contribution into $K^+\to
\pi^+$ transition} \label{1ac}\end{figure}
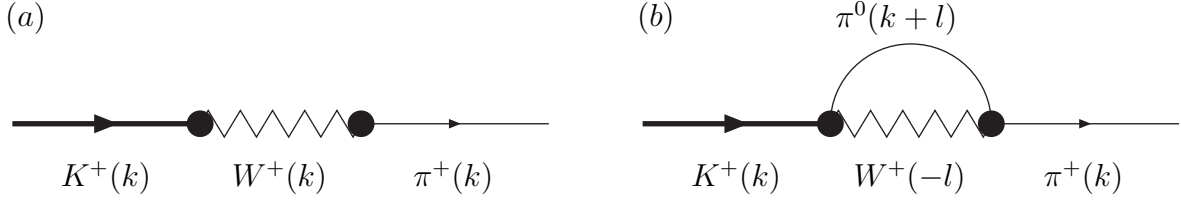

 Let us consider the $K^+\to \pi^+$ transition amplitude
\be\label{kp21}
 \langle\pi^+|-i\int dx^4dy^4
 J^\mu(x)D_{\mu\nu}^W(x-y)J^\nu(y)|K^+\rangle
 =i(2\pi)^{4}\delta^{4}(k-p)G_{\rm EW}\Sigma(k^{2}),
 \ee
 in the first order of the EW perturbation theory in the Fermi
 coupling constant (\ref{g})
 comparing two different W-boson field propagators,
 the accepted Lorentz (L)
 propagator (\ref{3-5-4b})
 and
 the  radiation (R) propagator (\ref{3-5-4}).
 These propagators give the expressions
 corresponding to the diagrams in Fig. \ref{1ac}
\bea\label{kp2}
  \Sigma(k^{2})&\to&
  \Sigma^R(k^{2})=2F^2_{\pi}{k^2}
  +2i\int \frac{d^4q{M_W^2}}{(2\pi)^4}
 \frac{k^2+(k_0+q_0)^2}{(-|\vec q|^2-M_W^2)[(k+q)^2
 -m^2_\pi+i\epsilon]}, \\
 \label{fkp2}
  \Sigma(k^{2})&\to&
  \Sigma^L(k^{2})=2F^2_{\pi}{k^2}
 +2i\int \frac{d^4q{M_W^2}}{(2\pi)^4}
 \frac{(2k_\mu+q_\mu)D^{L}_{\mu\nu}(-q)(2k_\nu+q_\nu)}{(k+q)^2
 -m^2_\pi+i\epsilon}.
 \eea
 The versions R and L coincide in the case of the axial contribution
 corresponding to the first diagram in Fig. \ref{1ac},
 and they both reduce to the static interaction contribution
 because
$$
k^\mu k^\nu D^F_{\mu\nu}(k)\equiv k^\mu k^\nu
D^R_{\mu\nu}(k)=\frac{k_0^2}{M^2_W}.
$$
 However, in the case of the vector contribution
 corresponding to the second diagram in Fig.  \ref{1ac}
 the radiation version differs from the  Lorentz gauge version
  (\ref{3-5-4b})\footnote{The Faddeev equivalence theorem \cite{f1}
 is not valid, because
  the vector current $J_\mu = K\partial_\mu \pi-\pi\partial_\mu K$
  becomes the vertex $\Gamma_\mu = K\partial_\mu D_\pi-D_\pi\partial_\mu K$,
   where one of fields
  is replaced by its propagator $\Box D_\pi =\delta (x)$, and
  $\partial_\mu \Gamma^\mu \not = 0$.}.

  In contrast to the Lorentz gauge version  (\ref{3-5-4b}),
  two radiation variable diagrams  in Fig. \ref{1ac} in  the
  rest kaon frame $k_\mu=(k_0,0,0,0)$
  are reduced to
  the static interaction contribution
 \be\label{kp1}
 i(2\pi)^{4}\delta^{4}(k-p)G_{\rm EW}\Sigma^R(k^{2})=
 \langle\pi^+|-i\int dx^4
 J_0(x)\frac{1}{\triangle-M_W^2}J_0(x)|K^+\rangle
 \ee
  with the normal ordering of the pion fields which are
  at their mass-shell\footnote{The second  integral in (\ref{kp2})
   with the term  $(k_0+q_0)^2$
   really does not depend on $k^2$, and it can be removed by the mass rotation.},
   so that
 \be\label{0Rkp1}
 \Sigma^R(k^{2}) = 2k^2{F^2_{\pi}}\left[1+
  \frac{M_{W}^{2}}{F^2_{\pi}(2\pi)^3}\int\frac{d^3l}{2E_\pi(\vec{l})}
  \frac{1}{M^2_W+\vec{l}^2}\right]
 \equiv 2k^2{F^2_{\pi}}g_8.
 \ee
Here $E_{\pi}(\vec{l})=\sqrt{m^{2}_{\pi}+\vec{l}^{2}}$ is the
 energy of $\pi$-meson and $g_8$ is
 the parameter of   the enhancement of the probability
 of the axial $K^+ \to \pi^+$ transition.
 The pion mass-shell justifies
 the application of the low-energy ChPT~\cite{fpp},
 where the summation of the chiral series can be
 considered here as
  the meson form factors
  \cite{bvp,belkov01,CP-enhancement}
  $\int\dfrac{d^3l}{2E_\pi(\vec{l})}\to$
  $\int\dfrac{d^3l f^V_K(-(\vec l)^2)f^V_\pi(-(\vec l)^2)}{2E_\pi(\vec{l})}$.

 Using
  the covariant perturbation theory \cite{pv}\ developed as the
 series $\underline{J}_\mu^k(\gamma
 \oplus\xi)=\underline{J}_\mu^k(\xi)+ F^2_\pi
 \partial_\mu \gamma^k +\gamma^if_{ijk}\underline{J}_\mu^j(\xi)+O(\gamma^2)$
 with respect to quantum fields $\gamma$ added to $\xi$
 as  the product $e^{i\gamma}e^{i\xi}\equiv e^{i(\gamma \oplus\xi)}$, one can see that the normal
  ordering $$<0|\gamma^i(x)\gamma^{i'}(y)|0>=\delta^{ii'}N(\vec
  z),~~~~
  N(\vec z)=\int \frac{d^3l e^{i\vec l\cdot (\vec z)}}{(2\pi)^32E_\pi(\vec l)},$$
  where $\vec z=\vec x-\vec y$,
  in the product of the currents $\underline{J}_\mu^k(\gamma
 \oplus\xi)$
   leads to an 
effective Lagrangian with the rule $\triangle T=1/2$ 
$$
 M_W^2\int d^3z g_8(|z|)\frac{e^{-M_W|\vec z|}}{4\pi|\vec z|}
 [\underline{J}_\mu^j(\xi(x))\underline{J}_\mu^{j'}(\xi(z+x))
(f_{ij1}+i f_{ij2})(f_{i'j'4}-if_{i'j'5})\delta^{ii'}+h.c] ,
 $$
 where
 $g_8(|z|)= [1+\sum\limits_{I\geq 1}c^I N^I(\vec z)]
 $
 is series
 over the multipaticle intermediate states (this sum is known as
 the Volkov superpropagator \cite{vp1,S}).
 In the  limit $M_W\to \infty$,
  in the lowest order with respect to $M_W$,  the
  dependence of $g_8(|\vec z|)$ and the currents on $\vec z$
 disappears in the integral of the type of
 $$M^2_W \int d^3z \dfrac{g_8(|\vec z|)e^{-M_W|\vec z|}}{4\pi|\vec z|}=
 \int_0^{\infty} drr e^{-r}g_8({r}/{M_W})\simeq g_8(0).$$
 In the next order, the amplitudes $K^0(\bar K^0)\to \pi^0$ arise.
  Finally, we get
  the effective  Lagrangians  \cite{kp}
 \be \label{ef1}\mathcal{L}_{(\Delta T=\frac{1}{2})}=
\frac{G_{F}}{\sqrt{2}}g_{8}(0)\cos\theta_{C}\sin\theta_{C}
\Big[(\underline{J}^1_{\mu}+i\underline{J}^2_{\mu})
(\underline{J}^4_{\mu}-i\underline{J}^5_{\mu})-
(\underline{J}^3_{\mu}+\frac{1}{\sqrt{3}}\underline{J}^8_{\mu})
(\underline{J}^6_{\mu}-i\underline{J}^7_{\mu})+h.c.\Big], \ee
\be\label{ef2} \mathcal{L}_{(\Delta T=\frac{3}{2})}=
\frac{G_{F}}{\sqrt{2}}\cos\theta_{C}\sin\theta_{C}
\Big[(\underline{J}^3_{\mu}+\frac{1}{\sqrt{3}}\underline{J}^8_{\mu})
(\underline{J}^6_{\mu}-i\underline{J}^7_{\mu})+h.c.\Big]. \ee

 This result shows that the enhancement can be explained
  by static vector interaction that
  increases the $K^+\to \pi^+$ transition
 by a factor of $g_8=g_8(0)$, and yields a new term describing the
  $K^0\to \pi^0$ transition proportional to $g_8-1$.

 This Lagrangian with the fit parameter $g_8=5$
 (i.e. $g_8\sin\theta_{C} \cos\theta_{C}\simeq 1$) 
 describes the nonleptonic decays in
 satisfactory agreement with experimental data 
 \cite{vp1,kp}.

\section*{Conclusions}

We have investigated the low-energy relations between the values of
resonance masses in the meson form factors, and the differential
radiative kaon decay $K^+\to \pi^+ e^+e^-(\mu^+\mu^-)$ rates
following from the ChPT~\cite{fpp,vp1,bvp}. We give non-trivial
predictions of muon/electron ratio and the effective form factor
slope, which are in agreement with the experimental data.

 The high  sensitivity of  these relations, the low energy status of
 ChPT, where they arise, and the universality of the enhancement  coupling
 constant $g_8$ for all kaon--pion weak transition amplitudes with
 the rule of selection $\triangle T =1/2$
 can be explained 
  by a  weak static
 interaction of massive vector bosons in the Hamiltonian
  approach to SM \cite{hpp}.

 The instantaneous character of
 weak static interaction in the Hamiltonian SM
  excludes  all retarded diagram contribution in the effective
   Chiral Perturbation
  Theory \cite{da98} that destruct
  the form factor structure of the kaon radiative decay rates.
  The enhancement of kaon--pion transition can  be considered as
a consequence of 
normal ordering of
 all pions in the instantaneous loop  on their mass-shells $p^2=m^2_\pi$.
 The static interaction mechanism of
  the enhancement of the $\triangle T=1/2$ transitions
 predicts   the coincidence of the meson form factor
 resonance parameters
 with the parameters of the radiation kaon decay rates $K^+ \to
 \pi^+e^{+}e^{-}(\mu^{+}\mu^{-})$ in
  satisfactory agreement with the experimental data \cite{7,ap99}.

  Therefore, the
  off-mass-shell kaon-pion transition in the radiation weak kaon decays
     can be a good probe of the weak static interactions.



\section*{Acknowledgements}

The authors are grateful to B.M. Barbashov, D.Yu. Bardin, A.Di
Giacomo, S.B. Gerasimov,  A.V. Efremov, V.D. Kekelidze, E.A. Kuraev,
V.B. Priezzhev, and the participants of the 5th Kaon Mini Workshop
(CERN, 12 December 2006) for fruitful discussions. The work was in
part supported by the Slovak Grant Agency for Sciences VEGA,
Gr.No.2/4099/26 and NA48/2 Project.

\appendix

\section{Appendix A: Calculation of $K^{+} \to \pi^{+} l^{+}l^{-}$ Decay Width}
\subsection{The Matrix Element}

\begin{figure}
\centering
\begin{minipage}[c]{0.45\hsize}
\begin{picture}(200,100)(0,0)
\Vertex(80,50){5} \Photon(80,50)(130,80){5}{4}
\ArrowLine(130,80)(200,100) \ArrowLine(200,60)(130,80)
\SetWidth{2.0} \ArrowLine(0,50)(80,50) \ArrowLine(80,50)(200,10)
\Text(35,30)[]{$K^{+}(k)$}
\Text(190,85)[]{$l^{-}(q_{-})$}\Text(190,50)[]{$l^{+}(q_{+})$}\Text(125,20)[]{$\pi^{+}(p)$}
\Text(95,85)[]{$\gamma^{*}(q)$}
\end{picture}
\end{minipage}
\caption{ $K^+\to\pi^+l^{-}l^{+}$
diagram.}\label{decay}\end{figure}
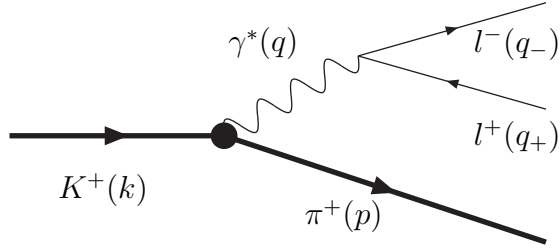

The matrix element for the process in Fig. \ref{decay} can be
obtained by Feynman rules:
\begin{equation}
i\mathcal{M}( K^{+} \to \pi^{+} l^{+}l^{-})  =
\bar{u}^{s}(q_{-})(-ie\gamma^{\mu})v^{s'}(q_{+}) \
i\frac{g_{\mu\nu}}{q^{2}} \  <\pi^{+}(p)| J_{em}{^\nu} |K^{+}(k)>,
\end{equation}
after inserting parametrization
\begin{equation}
i\mathcal{M}(K^{+} \to \pi^{+} l^{+}l^{-})  =
\bar{u}^{s}(q_{-})(-ie\gamma^{\mu})v^{s'}(q_{+}) \
i\frac{g_{\mu\nu}}{q^{2}} \ e F(kp) (k + p)^{\nu}.
\end{equation}
To obtain the square root of matrix element one has to sum over
spins

\bea \nonumber \sum_{s,s'} |\mathcal{M}|^2 & = & \frac{e^{4}|F(k
p)|^{2}}{q^{4}} Tr[q_{+} \! \! \! \! \slash \gamma_{\mu}\! \! \!
\! \slash q_{-}\! \! \! \! \slash \gamma_{\nu}\! \! \! \! \slash -
m^{2}_{l}\gamma_{\mu}\! \! \! \! \slash \gamma_{\nu}\! \! \! \!
\slash] (p+k)^{\mu}(p+k)^{\nu}
\\ \non
& = & \frac{4e^{4}|F(k p)|^{2}}{q^{4}}[
q_{+\mu}q_{-\nu}+q_{+\nu}q_{-\mu}-q_{+}q_{-}g_{\mu\nu}-
m^{2}_{l}g_{\mu\nu}](p+k)^{\mu}(p+k)^{\nu}
\\
& = & \frac{4e^{4}|F(kp)|^{2}}{q^{4}}[ 2
q_{+}(p+k).q_{-}(p+k)-q_{+}q_{-}(p+k)^{2}- m^{2}_{l}(p+k)^{2}].
\eea Next if we define

\be z=\frac{q^{2}}{M_{K}^{2}} =
\frac{2q_{+}q_{-}+2m^{2}_{l}}{M_{K}^{2}}; \quad
x=\frac{(p+q_{-})^{2}}{M_{K}^{2}} =
 \frac{m_{\pi}^{2}+2pq_{-}+m^{2}_{l}}{M_{K}^{2}}; \quad
R=\Big(\frac{m_{\pi}}{M_{K}}\Big)^{2};
r_l=\Big(\frac{m_{l}}{M_{K}}\Big)^{2}, \label{subst} \ee where
$l=e,\mu$ one obtains

\bea \nonumber \sum |\mathcal{M}|^2(z,x) & = &
\frac{2e^{4}F^{2}}{z^{2}}
\Big[\Big(2x+z-2-2r_l\Big)\Big(-2x-z+2R+2r_l\Big)+
z\Big(z-2-2R\Big)\Big],
\eea  where we used momentum conservation law and relations:
$$k=p+q_{-}+q_{+}=p+q,$$
$$q_{-}q_{+}=\frac{M_{K}^{2}}{2}(z-2r),$$
$$(k+p)^{2}=M_{K}^{2}(2+2R-z),$$
$$(k-p)^{2}=q^{2}=M_{K}^{2}z,$$
$$q_{-}(p+k)=\frac{M_{K}^{2}}{2}(2x+z-2R-2r),$$
$$q_{+}(p+k)=\frac{M_{K}^{2}}{2}(2+2r-z-2x),$$
$$q(p+k)=M_{K}^{2}(1-R).$$

\subsection{The decay rate}
The phase volume (in the frame of $\vec{k}=0$) is

\bea \non d\Phi & = &
\frac{d^{3}q_{+}d^{3}q_{-}d^{3}p}{2\epsilon_{+}2\epsilon_{-}2\epsilon}
\delta^{4}(k-p-q_{+}-q_{-}); \qquad \quad d^{4}q_{-}=d^{4}q, \\
\non & = &
\frac{|\vec{q}_{+}|\epsilon_{+}d\epsilon_{+}d\Omega_{+}}{2\epsilon_{+}}
d^{4}q
d^{4}p\delta^{4}(k-p-q)\delta((k-q)^{2}-m_{\pi}^{2})\delta((q-q_{+})^{2}-m_{l}^{2}),
\\ \non & = & \frac{|\vec{q}_{+}|d\epsilon_{+}d\Omega_{+}}{2.2M_{K}}\frac{1}{2}|\vec{q}|dq^{2}4\pi
\delta\Big(q^{2}-2q_{+}q\Big); \qquad d\Omega_{+}=2\pi
d\cos\theta_{+},
\\ & = &
\frac{d\epsilon_{+}}{8M_{K}}4\pi dq^{2}\frac{1}{2}2\pi =
\frac{dxdz}{16} 4\pi 2\pi M_{K}^{2}; \qquad
q_{+}q=q_{0}\epsilon_{+}-|\vec{q}||\vec{q}_{+}|\cos\theta_{+},
\eea and the decay width is

\bea \nonumber d\Gamma & = & \frac{1}{2M_{K}}
\frac{2e^{4}F^{2}}{z^{2}}
\Big[\Big(2x+z-2-2r_l\Big)\Big(-2x-z+2R+2r_l\Big)+
z\Big(z-2-2R\Big)\Big] \\ & \times & \frac{1}{(2\pi)^{5}}
\frac{dxdz}{4} \pi^{2} M_{K}^{2}. \label{width4} \eea

\bea \nonumber \frac{d\Gamma}{M_{K}dxdz} =
\frac{\alpha^{2}F^{2}}{8\pi z^{2}}
\Big[\Big(2x+z-2-2r_l\Big)\Big(-2x-z+2R+2r_l\Big)+
z\Big(z-2-2R\Big)\Big]. \label{width5} \eea After integration the
decay width is

\be \frac{d\Gamma}{M_{K} dz}  = \frac{\alpha^{2}}{6\pi z^{3}}
\lambda^{\frac{3}{2}}(1,z,R) F^{2}(z) (z+2r)\sqrt{1-\frac{4r}{z}},
\label{width3} \ee where

\be \lambda(1,z,R)=(z^{2}+1+R^{2}-2z-2Rz-2R).\ee


\begin{thebibliography}{}
%
\bibitem{fpp}
M. Volkov, V. Pervushin, Phys. Lett. {\bf B51}, 356 (1974).
%
\bibitem{vp1}
M. K. Volkov and V. N. Pervushin, Usp. Fiz. Nauk {\bf 120}, 363 (1976).\\
M. K. Volkov and V. N. Pervushin, ``Essentially Nonlinear Field
Theory, Dynamical Symmetry and Pion Physics'', Atomizdat, Moscow,
ed. D.I. Blokhintsev, 1979 (in Russian).
%
\bibitem{gs}
J. Gasser and H. Leutwyler, Ann. of Phys. {\bf 158}, 142 (1984).
%
\bibitem{bvp}
A.A. Bel'kov, Yu. L. Kalinovsky, V. N. Pervushin,
JINR-P2-85-107 (1985).\\
A.A. Bel'kov et al., Yad. Fis. {\bf 44}, 690 (1986).
%
\bibitem{ecker}
G. Ecker et al., Nucl. Phys. {\bf B291}, 692 (1987).
%
\bibitem{belkov01}
A.A. Bel'kov et al., Phys. Part. Nucl. {\bf 26}, 239 (1995).\\
A.A. Bel'kov, Phys. Part. Nucl. {\bf 36}, 509 (2005).
%
\bibitem{ecker-r}
G. Ecker, Prog. Part. Nucl. Phys. {\bf 36}, 71 (1996);
hep-ph/9511412.
%
%
\bibitem{da98}
G. D'Ambrosio et al., JHEP {\bf 9808}, 4 (1998).
%
\bibitem{05}
A.Z. Dubni\v{c}kov\'{a} et al., JINR E2-2006-80, Dubna, 2006;
hep-ph/0606005.
%
\bibitem{117}
A.Z. Dubni\v{c}kov\'{a} et al., hep-ph/0611175.
%
\bibitem{am}
G. Altarelli and L. Maiani, Phys. Lett. {\bf 52B}, 351 (1974).\\
M.K. Gaillard and B.W. Lee, Phys. Rev. Lett. {\bf 33}, 108 (1974).
%
\bibitem{va}
A.I. Vainshtein et al., Yad. Fiz. {\bf 24}, 820 (1976).\\
S.S. Gershtein and  M. Yu. Khlopov, JETP Lett. {\bf 23}, 338 (1976).
%

\bibitem{db}
D. Bardin and G. Passarino, ``The standard model in the making:
precision study of the electroweak interactions'', Clarendon,
Oxford, 1999.
%
\bibitem{hpp}
H.-P. Pavel  and V. N. Pervushin, Int. J. Mod. Phys. {\bf A14}, 2885
(1999). 
%
\bibitem{252} B. M. Barbashov et al., hep-th/0611252.

%
\bibitem {fp1}
L. Faddeev and V. Popov, Phys. Lett. {\bf B25}, 29 (1967).
%

%
\bibitem{kp}
Yu.L. Kalinovsky, V.N. Pervushin, Sov. J. Nucl. Phys., {\bf 29}, 225
(1979).
%
\bibitem{CP-enhancement}
A.A. Bel'kov  et al., Phys. Lett. {\bf B220}, 459 (1989).
%
\bibitem{5}
Yu.L. Kalinovsky et al., Sov. J. Nucl. Phys. {\bf 49}, 1059 (1989).\\
V.N. Pervushin, Nucl. Phys. B (Proc. Supp.) {\bf 15}, 197 (1990).
%
\bibitem{6}
V.N. Pervushin, Phys. Part. Nucl. {\bf 34}, 348 (2003).
%
\bibitem{pade} H. Lehmann, Phys. Lett. {\bf 41}, 529 (1972).\\
 J. Honerkamp, Nucl. Phys. {\bf B36}, 130 (1972).\\
 G. Ecker and J. Honerkamp,  Nucl. Phys. {\bf B52}, 211 (1973).\\
 M.K. Volkov and V.N. Pervushin,  Nuovo Cimento {\bf A27}, 277
 (1975).
%
\bibitem{7}
W.-M. Yao et al. (PDG), J. Phys. {\bf G33}, 1 (2006).
%
\bibitem{ap99}
R. Appel et al., Phys. Rev. Lett. {\bf 83}, 4482 (1999).
%
\bibitem{ad97}
S. Adler et al., Phys. Rev. Lett. {\bf 79}, 4756 (1997).
%
\bibitem{ma00}
H. Ma et al., Phys. Rev. Lett. {\bf 84}, 2580 (2000).
%
\bibitem{pa02}
H.K. Park et al., Phys. Rev. Lett. {\bf 88}, 111801 (2002). 
%
\bibitem{li99}
P. Lichard, Phys. Rev. {\bf D60}, 053007 (1999).

\bibitem{dir}
P.A.M. Dirac, Proc. Roy. Soc. {\bf A  114}, 243 (1927), Can. J.
Phys. {\bf 33}, 650 (1955). 
\bibitem{sch2}
J. Schwinger, Phys. Rev. {\bf 127}, 324 (1962).
%
\bibitem{f1}
L. Faddeev, T. M. F. {\bf 1}, 3 (1969).
%
\bibitem{ni}
N.P. Ilieva, N.S. Han, V.N. Pervushin, Sov. J. Nucl. Phys. {\bf 45},
1169 (1987).\\
N.S. Han, V.N. Pervushin, Mod. Phys. Lett. {\bf A2}, 367 (1987).\\
V.N. Pervushin, Nucl. Phys. B (Proc. Supp.) {\bf 15}, 197 (1990).
%
\bibitem{pv}
V.N. Pervushin, Theor. Math. Phys. {\bf 27}, 330 (1977).\\
D.I. Kazakov, V.N. Pervushin, S.V. Pushkin, {J. Phys. A. Math. Gen.}
{\bf 11}, 2093 (1978).
\bibitem{S}
M.K. Volkov, Ann. Phys. (N. Y.) {\bf 49}, 202 (1968).\\
C.J. Isham, Abdus Salam, and J. Strathdee, Phys. Rev. {\bf D3}, 1805
(1971).

\end{thebibliography}
\end{document}